# Pronounced Effect of pn-Junction Dimensionality on Tunnel Switch Threshold Shape

**Sapan Agarwal and Eli Yablonovitch**
Electrical Engineering & Computer Sciences Dept., University of California, Berkeley, 253 Cory Hall, Berkeley, CA, 94720

**Designing tunneling junctions with abrupt on-off characteristics and high current densities is critical for many different devices including backward diodes and tunneling field effect transistors (TFETs). It is possible to get a sharp, high conductance on/off transition by exploiting the sharp step in the density of states at band edges. The nature of the density of states, is strongly dependent on quantum dimensionality. To know the current/voltage curve requires us to specify both the n-side dimensionality and the p-side dimensionality of pn junctions. We find that a typical bulk 3d-3d tunneling pn junction has only a quadratic turn-on function, while a pn junction consisting of two overlapping quantum wells (2d-2d) would have the preferred step function response. We consider nine physically distinguishable possibilities: 3d-3d, 2d-2d$_{edge}$, 1d-1d$_{end}$, 2d-3d, 1d-2d, 0d-1d, 2d-2d$_{face}$, 1d-1d$_{edge}$ and 0d-0d. Thus we introduce the obligation to specify the dimensionality on either side of pn junctions.**

**Quantum confinement, or reduced dimensionality on each side of a pn junction has the added benefit of significantly increasing the tunnel conductance at the turn-on threshold.**

## Introduction

When designing tunneling junctions it is desired to achieve a very sharp turn on at low voltages. This is critical for backward diodes (1-4) and Tunneling Field Effect Transistors (TFETs) (5, 6). By using a sharp tunneling based switch it will be possible to significantly lower the voltage compared to conventional electronics.

To attain a sharp turn, the band edge energy filtering mechanism, or density of states overlap turn on, appears most promising. This mechanism is likely to provide high conductance, as well as sharp switching(7). This is illustrated in Fig. 1. If the conduction and valence band do not overlap, no current can flow. Once they do overlap, there is a path from filled valence band states to empty conduction band states for current to flow. Above threshold the turn on characteristic will be determined by the overlapping density of states in the conduction band and valence band. For example, we will find that in a typical 3d-3d bulk pn junction, the nature of the current-voltage (I-V) function beyond threshold is quadratic in the control voltage. A sharper density-of-states occurs if the dimensionality on either side of the pn junction is reduced. In addition, carrier confinement in the tunneling direction provides other benefits for increasing the on-state conductance.

Whenever specifying a pn junction it is also necessary to specify the dimensionalities of the respective p, and n regions. In Fig. 2 we show nine different possible pn junction dimensional combinations, and their corresponding tunnel I-V curves. In the following sections we analyze each of the dimensional combinations shown in Fig. 2: 1d-1d$_{end}$, 3d-3d, 2d-2d$_{edge}$, 0d-1d, 2d-3d, 1d-2d, 0d-0d, 1d-1d$_{edge}$, 2d-2d$_{face}$. We ask which are promising for adaptation into a TFET, or for a new type of Backward Diode?

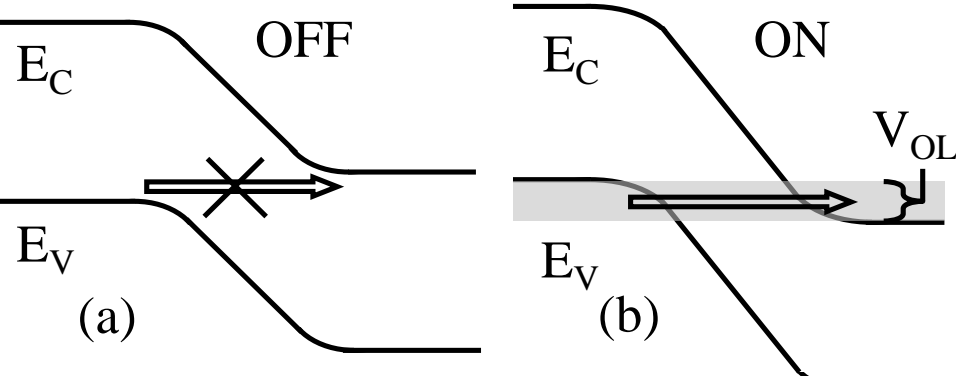

Fig. 1. (a) No current can flow when the bands do not overlap. (b) Once the bands overlap, current can flow. The band edges need to be very sharp, but density of states arising from dimensionality is also important.

### 1d-1d$_{end}$ Junction

A 1d-1d$_{end}$ pn junction, as shown in Fig. 2(a) describes tunneling within a nanowire (8) or carbon nanotube (9) junction. Tunneling is occurring from the valence band on the p-side to the conduction band on the n-side. For a transistor, the gate is not shown as there are many possible gate geometries. The corresponding band diagram is given by Fig. 3(a).

In analyzing all of the devices, we consider a direct gap semiconductor with a small bias. In particular we consider the regime near the band overlap turn-on where a small change in voltage ($k_bT/q$ or less) will result in a change of density of states overlap, but only a negligible change in the tunneling barrier thickness. Consequently, we assume that the tunneling probability is roughly a constant, $T$, and will not change

**Significance:**

**Minimizing power consumption is critical for modern electronics. Unfortunately, conventional electronics relies on thermal excitation of electrons over a barrier, necessitating an operating voltage many times larger than the thermal voltage, $k_bT/q$, to maintain a good on-off ratio. A new type of switch based on quantum mechanical tunneling has the potential to drastically lower the voltage and thus the power consumption. We show that the very shape of the I-V curve (linear vs quadratic vs step function, etc.) of these new switches strongly depends on the quantum dimensionality of the pn-junction. To realize an abrupt low voltage turn on, the dimensionality of a tunneling switch needs to be carefully engineered.**

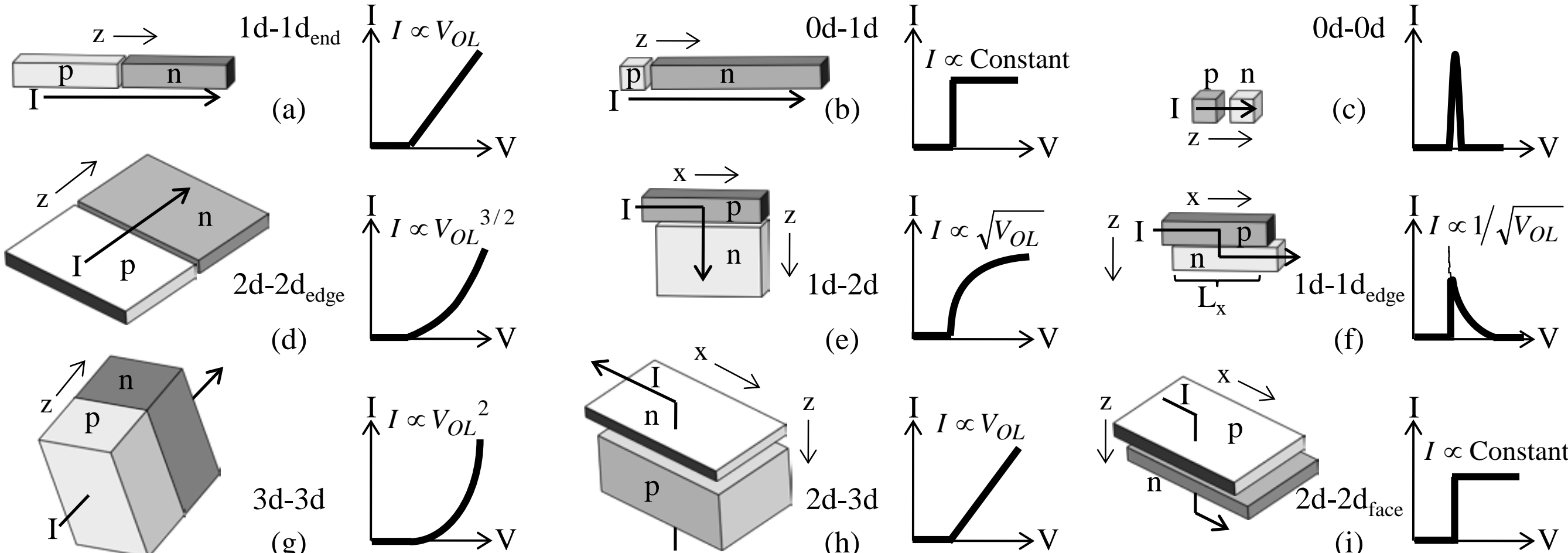

Fig. 2. We identify nine distinct dimensionality possibilities that can exist in pn junctions. Each of the different tunneling pn junction dimensionalities shown have different turn on characteristics as shown.

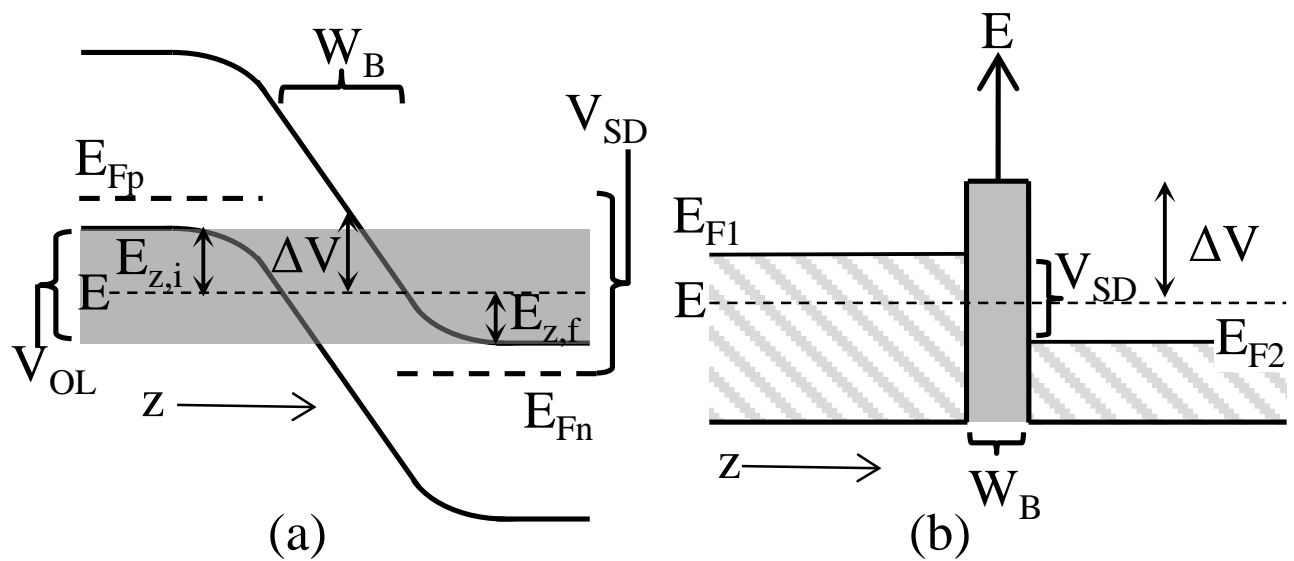

Fig. 3. (a) Energy band diagram for the tunnel pn junction showing that the relevant voltage is the overlap voltage and not the source drain voltage. (b) Energy band diagram for a typical 1d quantum of conductance showing that the relevant voltage is the source drain voltage.

significantly for small changes in the overlap voltage. (Nevertheless, $\top$ can be replaced with $\top$(voltage) in the following analyses if desired.) We also consider an energy averaged tunneling probability $\langle\top\rangle$. We will discuss the energy dependence $\top(E)$ later. The tunneling probability, $\top$, is the probability that an electron in a given mode tunnels through the barrier and end up on the other side. It is often specified by the WKB approximation: $\top = \exp\left(\int kdz\right)$, where $k$ is the imaginary wave vector in the barrier.

We also define $V_{OL}=qE_{OL}$ to be the overlap voltage between the conduction and valence bands as shown in Fig. 3(a). In order to keep the analysis as simple and general as possible we will use the band overlap voltage, $V_{OL}$ in all of the analyses instead of other applied voltages. Our goal is to focus on the quantum and density of states effects rather than electrostatic effects, as is done elsewhere(10).

The 1d-1d$_{end}$ current can be derived as an adaptation of the normal quantum of conductance, $2q^2/h$, approach shown in Fig. 3(b). The current flow is controlled by the difference in the Fermi levels: $E_{F1}$-$E_{F2}$≡$qV_{SD}$. $V_{SD}$ is the voltage applied to the ends of the 1d wire. The current is $I=(2q^2/h)\times V_{SD}\times\langle\top\rangle$.

Now to properly consider the transition from conduction band to valence band, consider the band diagram given in Fig. 3(a). Initially consider the valence band on the p-side of the junction to completely full and the conduction band on the n-side to be completely empty. This requires $V_{SD}>k_bT/q$ and $V_{SD}>V_{OL}$.

As shown in Fig. 3(a), the band edges determine the energy levels that can contribute to the current. Unlike a single band 1d conductor, the overlap voltage $V_{OL}$ determines the amount of current that can flow. Consequently, it is $V_{OL}$ and not $V_{SD}$ that controls the current:

$$I_{1d-1d} = \frac{2q^2}{h}\times V_{OL}\times\langle\top\rangle \qquad \textbf{[1]}$$

***$V_{SD}$ < 4$k_bT$ Limit:***

We can also consider the opposite limit where $V_{SD}<4k_bT/q$. The tunneling proceeds from from filled to empty states represented by $f_v(1 - f_c)$ minus the reverse process $f_c(1 - f_v)$ netting out to $f_c - f_v$, where:

$$f_{c,v} = \frac{1}{e^{(E-E_{Fc,v})/k_bT}+1} \qquad \textbf{[2]}$$

The tunneling current is diminished by $f_c$ - $f_v$:

$$I_{1d-1d} = \frac{2q}{h}\int_0^{qV_{OL}}\times\langle\top\rangle\times(f_c - f_v)\times dE \qquad \textbf{[3]}$$

This what one would find following the Landauer approach for a single mode. In this small bias regime everything of interest occurs within ~$k_bT$ of energy. Consequently, we can Taylor expand $f_c$ - $f_v$:

$$f_c - f_v \approx \frac{(E_{Fc}-E_{Fv})}{4k_bT} \approx \frac{qV_{SD}}{4k_bT} \qquad \textbf{[4]}$$

Thus the ultimate effect of the small differential Fermi occupation factors is to multiply the large bias current by the factor $qV_{SD}/4k_bT$. We can therefore write the current and conductance for small source drain biases:

$$I_{1d-1d} = \int_0^{qV_{OL}}\frac{2q}{h}\times\langle\top\rangle\times\frac{qV_{SD}}{4k_bT}\times dE \qquad \textbf{[5a]}$$

$$I_{1d-1d} = \frac{2q^2}{h} \times \langle T \rangle \times V_{OL} \times \frac{qV_{SD}}{4k_b T} \quad \textbf{[5b]}$$

$$G_{1d-1d} = \frac{2q^2}{h} \times \langle T \rangle \times \frac{qV_{OL}}{4k_b T} \quad \textbf{[6]}$$

We consider conductance rather than current, as the speed of a low voltage device is limited by its RC time and not by its current density. This is true for all of the following devices to be considered in next sections as well, but for brevity we will consider only the opposite limit where $f_c$ - $f_v$=1. The following sections provide the I-V curve for 3d-3d, 2d-2d$_{edge}$, 2d-3d, 1d-2d, 0d-1d, 2d-2d$_{face}$, 1d-1d$_{edge}$ and 0d-0d.

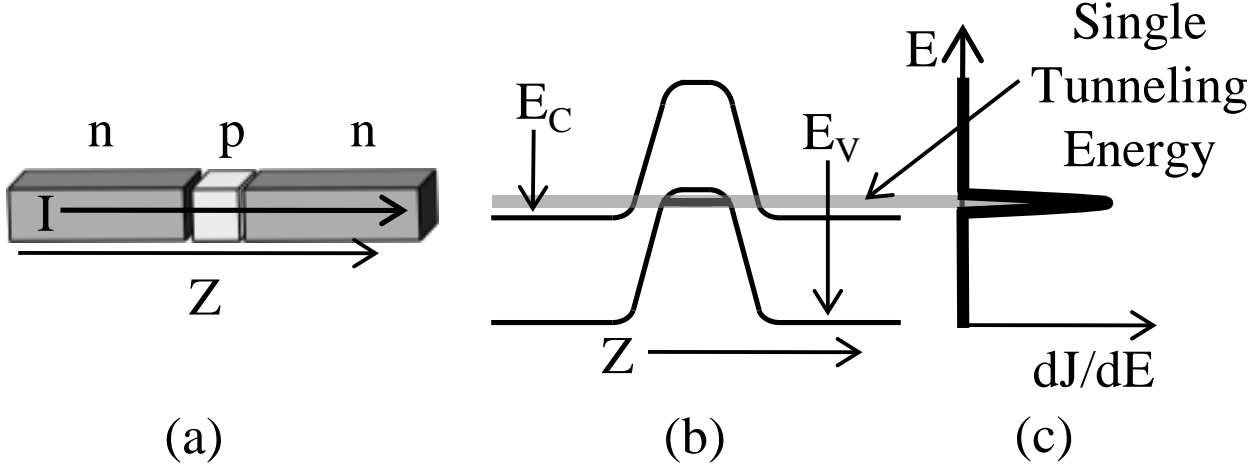

Fig. 4: (a) A 0d-1d junction converted into a more realistic 1d single electron transistor (SET) structure. The left nanowire can also be a p-type wire. (b) Band diagram corresponding to the SET. (c) All the current is concentrated around a single energy, which allows for a small overlap voltage $V_{OL}$, and thus a sharp turn on.

### 3d-3d Bulk Junction

A 3d-3d junction simply means a pn junction or heterojunction where there is a bulk semiconductor on either side of the sample as shown in Fig. 2(g). To derive the current we need to sum the 1d-1d$_{end}$ result over the number of transverse modes:

$$\partial I = N_{\perp\, states} \times \frac{2q}{h} \times \langle T \rangle \times \partial E \quad \textbf{[7]}$$

Integrating Eq. **7** gives:

$$I_{3d-3d} = \frac{1}{2}\left( \frac{Am^*}{2\pi\hbar^2} \times \frac{qV_{OL}}{2} \right) \times \frac{2q^2}{h} V_{OL} \times \langle T \rangle \quad \textbf{[8]}$$

=No. of 2d Channels × 1d Conductance

Where A is the area of the tunneling junction.

### 2d-2d$_{edge}$ Junction

A 2d-2d$_{edge}$ junction is shown in Fig. 2(d). The derivation of the current is almost identical to the 3d-3d case, except that instead of having a 2d array of 1d channels we now have a 1d array of 1d channels. Therefore the current is:

$$I_{2d-2d,edge} = \frac{2}{3}\left( \frac{L_x \sqrt{m^*}}{\pi\hbar} \times \sqrt{qV_{OL}} \right) \times \left( \frac{2q^2}{h} \times V_{OL} \times \langle T \rangle \right) \quad \textbf{[9]}$$

=No. of 1d Channels × 1d Conductance

Where $L_x$ is the length of the junction.

### 0d-1d Junction

A 0d to 1d junction represents tunneling from a quantum dot to a nanowire as shown in Fig. 2(b) or more realisitically in Fig. 4(a). In Fig. 2(b), we will assume that there is an electron in the quantum dot and find the rate at which it escapes into the end of a 1d wire. We analyze this junction as building block for the 2d-3d and 1d-2d junctions. To build a real 0d-1d device, we also need to electrically contact the quantum dot. This becomes a single electron transistor (SET) as shown in Fig. 4.

The 0d-1d contact current can be given by the rate at which an electron escapes from the quantum dot into a nanowire. The particle is oscillating back and forth in its well and it attempts to tunnel out on each round trip oscillation (11). If the dot has a length of $L_z$ along the tunneling direction, the electron will travel a distance of $2L_z$ between tunneling attempts. Its momentum is given by $p_z=mv_z=\hbar k_z$ where $k_z=\pi/L_z$ in the ground state. Using $E_z = \hbar^2 k_z^2/2m$, the time between tunneling attempts is $\tau=2L_z/v_z=h/2E_z$. The tunneling rate per second is $R=(1/\tau)\times\langle T \rangle$. This can be converted to a current by multiplying by the electron charge, and a factor 2 for spin to give:

$$I = \frac{4q}{h} \times E_z \times \langle T \rangle \quad \textbf{[10]}$$

To include coupling into the dot, we add a second nanowire to supply current, as shown in Fig. 4 and form a "single electron transistor" (12). Unlike a conventional SET, we want the current to be high enough and the dot be large enough to eliminate any coulomb blockade effects, which could interfere with switching action. Since the tunneling event out of the dot follows sequentially after tunneling in, the current is cut in half:

$$I_{0d\text{-}1d} = \frac{2q}{h} \times E_z \times \langle T \rangle \quad \textbf{[11]}$$

As seen in Fig. 4(c), the tunneling occurs at a single energy and will result in a step function turn on, once the bands overlap. The current will remain constant as long as the dot level overlaps with the bands in both the initial and final wire. This is one of the key benefits of quantum confinement. The current density is concentrated in a narrow energy range which allows for a sharper I-V curve. This can be contrasted with the 1d-1d$_{end}$ case, Eq. **1**, where the current flows over the entire energy range corresponding to $qV_{OL}$. The width of the 0d-1d energy range will be given by the broadening of the energy level in the quantum dot. This broadening can be extrinsically caused by any inhomogeneities in the lattice such as defects, dopants, or phonons. Even without these effects, simply coupling to the dot to the nanowires causes a significant amount of broadening. Each contact will broaden the level by $\gamma_0$ for a total broadening of $2\gamma_0$ (13), where:

$$\gamma_0 = \frac{\hbar}{\tau} = \frac{1}{\pi} \times E_z \times \langle T \rangle \quad \textbf{[12]}$$

In the limit that $\langle T \rangle \rightarrow 1$, the 0d-1d case degenerates to the 1d-1d$_{end}$ case. Nonetheless, in a realistic situation, $\langle T \rangle << 1$, and so we can use quantum confinement to concentrate the current at a single energy, with a significantly sharper step-function I-V curve.

### 2d-3d Junction

A 2d-3d tunneling junction is typical in vertical tunneling junctions where the tunneling occurs from the bulk to a thin confined layer as shown in Fig. 2(h). The thin layer can either be a thin inversion layer or a physically separate material (14-16). To find the 2d-3d current, we simply multiply the 0d-1d result, Eq. **10**, by the number of 2d channels to get a current of:

$$I_{2d-3d} = \text{No. of 2d channels} \times \text{0d - 1d contact current}$$

$$I_{2d-3d} = \left(\frac{Am}{2\pi\hbar^2}\times\frac{qV_{OL}}{2}\right)\times\left(\frac{4q}{h}\times E_z\times\langle\top\rangle\right) \quad \textbf{[13]}$$

Here, $E_z$ is the confinement energy of the 2d layer. Compared to the bulk 3d-3d case, confining one side of the junction resulted in the replacement of $qV_{OL}$ with $4E_z$.

Current can flow in along the x-direction as shown in Fig. 2(h). Other methods such as tunneling into the quantum well can also be considered for making electrical contact.

### 1d-2d Junction

A 1d-2d junction describes tunneling between the edge of a nanowire and a 2d sheet as shown in Fig. 2(e). The derivation for this case is almost identical to the 2d-3d case. The only difference is that instead of a 2d array of 1d tunneling, we now have a 1d array of 1d tunneling. Thus the current is:

$$I_{1d\text{-}2d} = \text{no. of 1d channels} \times \text{0d - 1d contact current}$$

$$I_{1d-2d} = \left(\frac{L_x}{\pi\hbar}\times\sqrt{qm^*V_{OL}}\right)\times\left(\frac{4q}{h}\times E_z\times\langle\top\rangle\right) \quad \textbf{[14]}$$

Comparing to the 2d-2d edge overlap formula, confining one side of the junction resulted in the replacement of $qV_{OL}$ with $3E_z$.

### Fermi's Golden Rule Derivation

The current for all of the dimensionalities can be derived in a manner different from above, using the transfer Hamiltonian method (17-20). We do this now as an alternative to employing the more modern channel conductance approach that we have been using. The transfer Hamiltonian method is just an application of Fermi's golden rule:

$$J = 2q\times\frac{2\pi}{\hbar}\sum_{k_i,k_f}\left|M_{fi}\right|^2\delta(E_i-E_f)(f_c-f_v) \quad \textbf{[15]}$$

The calculation of the matrix element $M_{fi}$ is in done in SI Appendix A and Ref. (19) and is given by:

$$M_{fi} = \frac{\hbar^2}{2m}\sqrt{\frac{k_{z,f}\,k_{z,i}}{L_{z,f}\,L_{z,i}}}\times\sqrt{\top}\times\delta_{k_{x,i},k_{x,f}}\,\delta_{k_{y,i},k_{y,f}} \quad \textbf{[16]}$$

In this equation, $k_{\alpha,i}$ and $k_{\alpha,f}$ are the wave-vectors in the initial and final states respectively, and $\alpha$ is the Cartesian index. $L_{z,i}$ and $L_{z,f}$ are the lengths of the initial and final sides of the junction, along the tunneling direction. This method is convenient in some of the reduced dimensionality cases, as discussed in SI Appendix B.

When quantum confinement is employed in the tunneling direction, two effects will result in a larger matrix element and thus a higher conductance. 1) The wave vector in the tunneling direction, $k_z$ will be larger, leading to increased velocity, and a higher tunnel attempt rate. 2) In the quantum confinement direction, $L_z$ will be shorter. By shrinking the confinement region, a greater percentage of the electron density penetrates the barrier and thus the tunneling wave-function overlap increases.

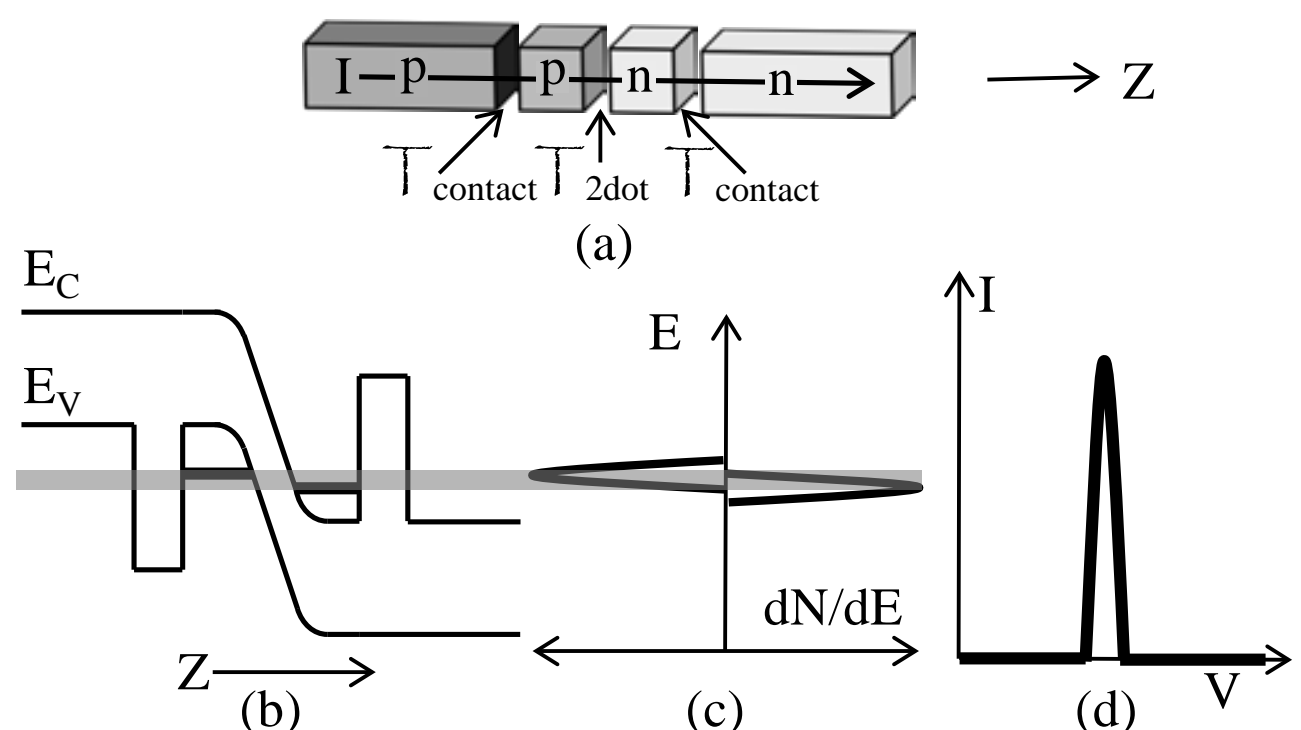

Fig. 5: The properties of a 0d-0d junction that is coupled to nanowire contacts are shown. (a) Schematic representation of the junction (b) Band diagram of the junction. (c) Tunneling only occurs at a single energy when the density of states in each dot is aligned. (d) The I-V curve resembles a delta function when the levels align.

### 0d-0d Junction

This case represents tunneling from a filled valence band quantum dot to an empty conduction band quantum dot. It is schematically represented in Fig. 2(c). In order to create a meaningful device, the quantum dots need to be coupled to contacts to pass current in and out of the device as shown in Fig. 5.

Current will only flow when the confined energy levels in each dot are aligned. This can be seen from Fig. 5(d). This results in an I-V curve that resembles a delta-function as shown in Fig. 5(e). We can estimate the peak current by considering the coupling strength between each dot and its contact as well as the coupling between dots. For simplicity, we will assume that the dots and contacts are symmetric. The coupling strength or broadening due to each contact, $\gamma_0$, is given by Eq. **12** where $\top$ is replaced by $\top_{contact}$ which represents the tunneling probability between the contact and a dot.

The matrix element between the initial state i on one dot and final state f on the other dot is given by Eq. **16**. Since we have a single level in each dot, we can simplify the matrix element by using $k_z=\pi/L_z$ and $E_z = \hbar^2 k_z^2/2m^*$:

$$\left|M_{f\,i,0d-0d}\right| = \frac{1}{\pi}\sqrt{E_{z,i}\times E_{z,f}\times\langle\top_{2dot}\rangle} \quad \textbf{[17]}$$

$\top_{2dot}$ is the single barrier tunneling probability between the two dots.

For a single level, the following form of Fermi's Golden Rule provides the rate of current flow:

$$I = 2q \times R_{fi} = 2q \times \frac{2\pi}{\hbar} |M_{fi}|^2 \frac{dN}{dE} \qquad \textbf{[18]}$$

where the 2 is present to allow for both electron spin polarizations and $dN/dE$ is the density of final states. If we consider a simple Lorentzian lineshape model for a 0d level, the peak density of states is: $dN/dE = 2/\pi\gamma = 2/(E_z \times \top_{contact})$. Since the 0d-0d system has contacts on both sides, the system broadening is doubled, which reduces dN/dE by a factor 2 becoming $dN/dE = 1/(E_z \times \top_{contact})$. Plugging in the peak $dN/dE$ and the matrix element, Eq. **17**, into Eq. **18** gives:

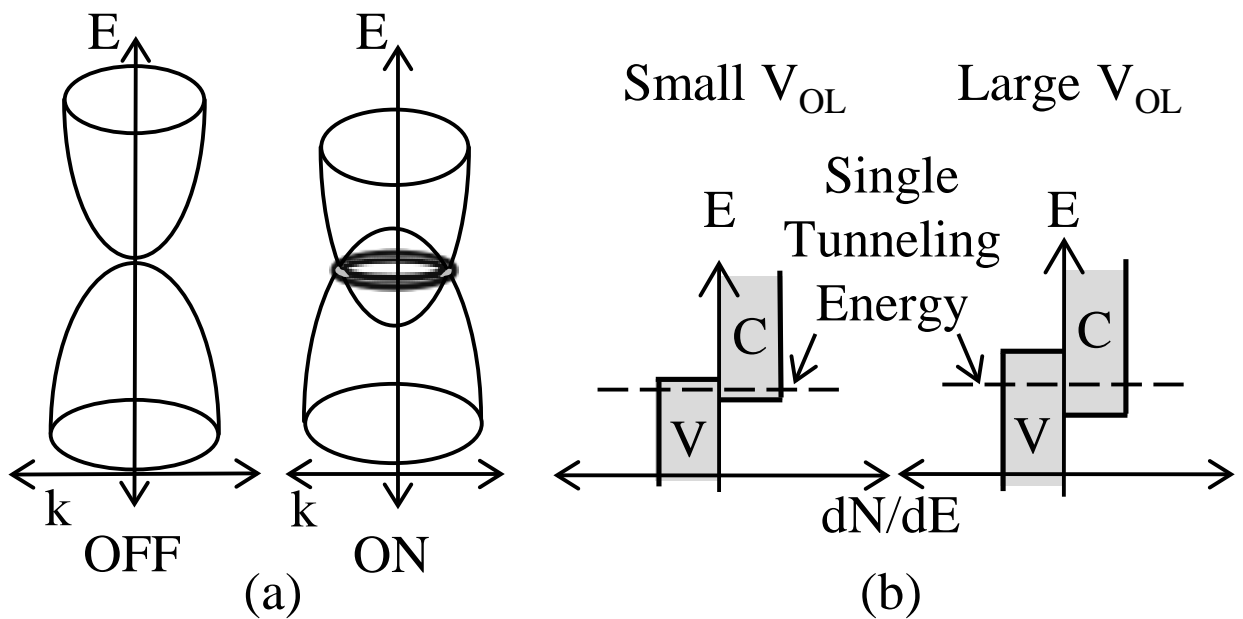

Fig. 6: Various characteristics of a 2d-2d$_{face}$ junction. (a) There is only a single tunneling energy because of the simultaneous conservation of energy and momentum. The energy versus wave vector paraboloids on each side of the junction only intersect at a single energy. (b) Even though the overlap of the density of states increases with increasing overlap voltage, there is only a single energy, indicated by the dotted line, at which the electrons tunnel.

$$I_{peak} = 2q \frac{2\pi}{\hbar} \left[ \frac{1}{\pi^2} E_z \times E_z \times \langle \top_{2\mathrm{dot}} \rangle \right] \frac{1}{E_z \times \langle \top_{\mathrm{contact}} \rangle} \qquad \textbf{[19]}$$

$$I_{peak} = \frac{8q}{h} E_z \times \frac{\langle \top_{2\mathrm{dot}} \rangle}{\langle \top_{\mathrm{contact}} \rangle} \qquad \textbf{[20]}$$

If the perturbation matrix element is stronger than the broadening, $|M_{fi}| > \gamma$, the two levels will strongly couple, leading to level repulsion preventing the desired conductance switching when levels align. Consequently, for Fermi's Golden Rule to be valid we need $|M_{fi}|/\gamma < 1$. Avoiding level repulsion provides a subsidiary requirement:

$$\frac{|\mathrm{M_{fi}}|}{\gamma} = \frac{\sqrt{\langle \top_{2\mathrm{dot}} \rangle}}{\langle \top_{\mathrm{contact}} \rangle} < 1, \text{ or } \langle \top_{2\mathrm{dot}} \rangle < \langle \top \mathrm{contact} \rangle^2 \qquad \textbf{[21]}$$

This limits the current to:

$$I_{peak} < \frac{8q}{h} E_z \langle \top_{contact} \rangle \qquad \textbf{[22]}$$

The width of the tunneling peak is given by the broadening of the confined level, $2\gamma_0/q$. Additional broadening mechanisms such as electron-phonon interactions can further broaden the I-V curve and reduce the peak current by smearing out the levels and reducing the coupling strength between the dots. As with the 0d-1d case, in the limit that $\langle \top \rangle \rightarrow 1$, the 0d-0d case will degenerate to the 1d-1d$_{end}$ case with a perfect quantum of conductance: $I = 2q^2/h \times V_{OL}$. However, in a realistic situation $\langle \top_{\mathrm{contact}} \rangle << 1$, sharpening up the linewidth γ, and concentrating the current at a single energy and voltage, allowing for abrupt switching.

### 2d-2d$_{face}$ Junction

A 2d-2d$_{face}$ junction describes tunneling from one quantum well to another through the face of the quantum well. This can be seen in resonant interband tunnel diodes (21-23) and the electon hole bilayer TFET(24, 25). The junction is schematically represented in Fig. 2(i). This is one of the most interesting cases as it is close to a step function I-V turn-on curve.

The step function turn-on can be derived by considering the conservation of transverse momentum and total energy. This depicted in Fig. 6(a). The lower paraboloid represents all of the available states in k-space on the p side of the junction and the upper paraboloid represents the available k-space states on the n side of the junction. In order for current to flow the initial and final energy, and wave-vector k, must be the same and so the paraboloids must overlap. However, as seen in the right part of Fig. 6(a), they can only overlap at a single energy while conserving energy and momentum. Within the joint density of states between valence and conduction band, only a single energy is responsible for the tunneling, regardless of the overlap as seen in Fig. 6(b). Since the joint density of states is constant in 2 dimensions, the tunnel current will also be constant, leading to a step function I-V curve.

The current can be computed by using Fermi's golden rule. Due to the conservation of transverse momentum, every initial state is coupled to only one final state. Plugging in the 0d-0d matrix element, Eq. **17** into Fermi's golden rule, Eq. **15**, and converting the sums over transverse k to an integral over transverse energy $E_t$, and constraining $E_f = E_i$ gives:

$$I = 2q \times \frac{Am}{\pi^2 \hbar^3} \int E_{z,i} \times E_{z,f} \times \langle \top \rangle \times \delta(E_i - E_f) dE_t \qquad \textbf{[23]}$$

where $E_{z,i}$ and $E_{z,f}$ are the confinement energies in the 2d quantum well. An additional factor of ½ appears when evaluating the integral owing to the sum of conduction and valence band transverse energy, $E_i - E_f = 2E_t - qV_{OL}$ :

$$I_{2d-2d,face} = \frac{qmA}{\pi^2 \hbar^3} \times E_{z,i} \times E_{z,f} \times \langle \top \rangle \qquad \textbf{[24]}$$

The main change in going from 3d-3d to 3d-2d is that the overlap energy $qV_{OL}$ became the quantum confinement energy, $E_z$. Likewise, in going from the 3d-2d to 2d-2d$_{face}$ the other overlap energy $qV_{OL}$ also became $E_z$. Thus for each confined side of the junction the relevant energy changes from the overlap energy to the confinement energy. In practice $E_z$ can be much larger than $qV_{OL}$, providing the 2d-2d$_{face}$ case with a significant current boost.

### 1d-1d$_{edge}$ Junction

A 1d-1d$_{edge}$ junction represents two nanowires overlapping each other along the edge as shown in Fig. 2(f). Similar to the 2d-2d$_{face}$ junction the current can be found using Fermi's golden rule, Eq. **15**. The resulting current is:

$$I_{1d-1d,\, edge} = 2\frac{qL_x}{\pi^2\hbar^2} E_{z,i} \times E_{Z,f} \times \sqrt{\frac{m}{qV_{OL}}} \times \langle \top \rangle \qquad \textbf{[25]}$$

As in the 2d-2d$_{face}$ case the tunneling only occurs at a single energy due to the conservation of momentum and energy. Since we are now dealing with 1d nanowires, the number of transverse states follows a 1d density of states which follows a $1/\sqrt{V_{OL}}$ dependence. This predicts a step turn on followed by a reciprocal square root decrease. This seemingly implies that the initial conductance will be infinite. However, the contact series resistance and various broadening mechanisms will limit the peak conductance.

**Energy Dependent Tunneling Probability**

Now that we have analyzed the different dimensionalities, we consider some corrections that occur near turn on. The tunneling probability has a significant energy dependence at low energies where there is a prefactor to the WKB exponential. At energies small relative to the barrier height, the wave function, $\psi$, resembles the infinite barrier case, in which $\psi$=0 at the barrier. Therefore, at low energy the tunneling probability approaches zero.

For rectangular wells, the tunneling probability can be found from rectangular well wavefunctions when evaluating the matrix element Eq. **16** as is done in SI Appendix A. At small energies, much lower than the barrier height, we get:

$$\top \approx \frac{16\sqrt{E_{z,i}E_{z,f}}}{\Delta V}\exp\left(-2\kappa W_B\right) \qquad \textbf{[26]}$$

The initial and final kinetic energy are given by $E_{z,i}$ and $E_{z,f}$ respectively. The barrier height minus the tunneling energy, $E$, is given by $\Delta V$ as shown in Fig 3. The barrier width is $W_B$. The imaginary wavevector in the tunneling barrier is given by $\kappa = \sqrt{2m\Delta V}/\hbar$.

Thus we see that the WKB exponential should also include an energy dependent prefactor. As the energy goes to zero, the tunneling probability goes to zero. This influences the key voltage dependence. When there is no confinement in the tunneling direction, such as the 1d-1d$_{end}$, 2d-2d$_{edge}$ and 3d-3d cases, the available kinetic energy, $E_z$, is limited by $V_{OL}$ and the prefactor provides additional voltage dependence. Using an energy averaged tunneling probability as we have previously done will still capture this voltage dependence. The prefactor will saturate near unity for a large $E_z$ and therefore large $V_{OL}$. Consequently, the initial turn on current will be lowered until the prefactor saturates.

Conversely, confinement in the 0d-0d, 1d-1d$_{edge}$ and 2d-2d$_{face}$ cases, fixes $E_{z,i}$ and $E_{z,f}$, which can be larger than $\Delta V$, making the prefactor in Eq. **26** saturate at unity, as discussed in SI Appendix A. In the next section we introduce additional level broadening mechanisms that smear the thresholds.

**Tradeoff between Current, Device Size, and Level Broadening**

When a level on the p-side of a junction interacts with a level on the n-side of the junction it is possible for the two levels to interact strongly and repel each other. For large contact regions leading to the tunnel junction, individual wave functions are spread out over a large normalization length, guaranteeing that individual level repulsion matrix elements are negligible compared to any residual broadening.

In contrast, the 0d-0d, 1d-1d$_{edge}$ and 2d-2d$_{face}$ cases are confined along the tunneling direction, restricting the normalization length. This means that the tunnel interaction matrix element, $|M_{fi}|$, can take on a large finite value, promoting level repulsion and preventing On/Off switching action. To avoid this and wash out the level repulsion, the level broadening, $\gamma$, needs to be greater than the level repulsion matrix element:

$$\gamma > \left|M_{fi}\right| = \frac{1}{\pi}\sqrt{E_{z,i} \times E_{z,f} \times \langle \top \rangle} \qquad \textbf{[27]}$$

This is the same as Eq. **21**, but applies to the 1d-1d$_{edge}$ and 2d-2d$_{face}$ cases in addition to 0d-0d.

The broadening $\gamma$ is typically caused by coupling to the contacts or by various scattering mechanisms. While this level broadening smears out the sharp I-V curve of the 0d-0d, 1d-1d$_{edge}$ and 2d-2d$_{face}$ junctions, it is required to prevent level repulsion. Consequently, it is necessary to engineer the tradeoff between the smearing and the tunnel conductance.

Another major broadening limit occurs for the 1d-2d, 2d-3d, 1d-1d$_{edge}$ and 2d-2d$_{face}$ cases when the overlap length, $L_x$, is reduced. Consider the 1d-1d$_{edge}$ case shown in Fig. 2(f). The electrons in the overlap region can tunnel, but they also need to escape from the overlap region into the contacts. This leads to a level broadening. The broadening is given by the energy of an electron confined by the overlap length $L_x$. The escape time, $\tau$, for an electron in an overlap length $L_x$ is $L_x$/v, and v is obtained from the quantum confinement velocity $\hbar k_x / m = \hbar\pi / mL_x$, within the confinement length $L_x$. The escape rate $1/\tau = v/L_x = \hbar\pi / mL_x^2$ leads to an "escape time" energy broadening:

$$\gamma \approx \frac{\hbar^2\pi}{mL_x^2} = \frac{2}{\pi}E_x \qquad \textbf{[28]}$$

where $E_x$ is confinement energy along the overlap region. For all the cases, 1d-1d$_{edge}$, 1d-2d, 2d-3d and 2d-2d$_{face}$,where overlap is important, the onset of the I-V threshold will be smeared by Eq. **28**.

In addition to contact broadening, and "escape time" broadening, there are a wide variety of additional broadening mechanisms that can smear out the initial turn on: phonon bandtails, phonon side-bands, Coulomb Blockade threshold shift, charge noise, etc. There are also device problems such spatial inhomogeneity, doping-induced inhomogeneity and trap states.

**Comparing the Different Dimensionalities**

Now that we have considered many different tunneling junction geometries, we've plotted a comparison of the different cases in Fig. 7. To plot the figures we used a reasonable tunneling probability of 1%. We assumed confinement energies of 130 meV, an effective mass of $0.1m_o$, overlap lengths of 20 nm, and a barrier height of $\Delta V$=100 meV. For the 0d-0d case we assumed $\top_{2dot} = ¼ \times \top_{contact}^2$ to satisfy Eq. **21**. We also consider four intrinsic broadening mechanisms that will limit the initial turn on, as indicated by the dotted lines. Using the above constants, the broadening mechanisms and the affected dimensionalities are summarized below:

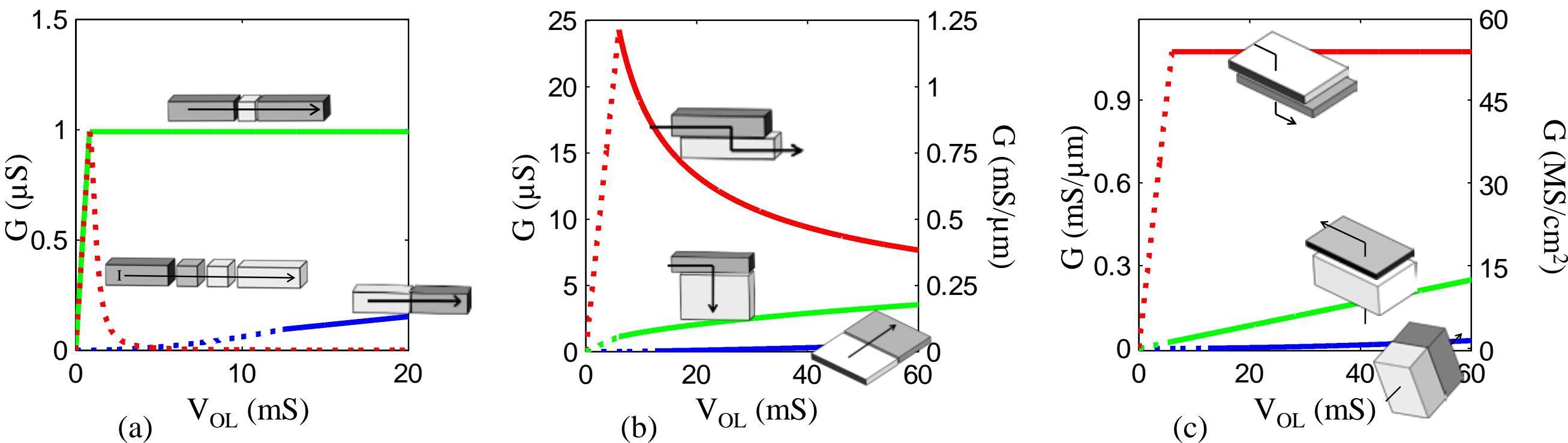

Fig. 7: The conductance curves for the different dimensionalities are plotted using the following parameters: $\langle\top\rangle$=1%, $E_c$=130meV, $L_x$=20nm, $m^*=0.1m_e$ and $\Delta V$=100meV. The dotted lines represent the initial broadened turn on where the lineshape is uncertain. (a) The 1d-1d$_{end}$, 0d-1d and 0d-0d cases are plotted. (b) The 2d-2d$_{edge}$, 1d-2d and 1d-1d$_{edge}$ cases are plotted. (c) The 3d-3d, 2d-3d and 2d-2d$_{face}$ cases are plotted. For the 0d-0d case the entire line-shape is linked to the broadening and is thus unknown, but the calculated width and height are still represented in the figure.

- Escape time broadening: Eq. **28**
  - 1d-2d, 2d-3d, 1d-1d$_{edge}$, 2d-2d$_{face}$, ------------ $\gamma$=6.0 meV
- Contact Broadening: Eq. **12**
  - 0d-0d, and 0d-1d -------------------------------- $\gamma$=0.8 meV
- Matrix Element Level Repulsion: Eq. **27**
  - 1d-1d$_{edge}$ and 2d-2d$_{face}$ -------------------------- $\gamma$=4.1 meV
  - 0d-0d ---------------------------------------------- $\gamma$=0.2 meV
- Weak WKB Tunneling Prefactor: Eq. **26**
  - 1d-1d$_{end}$, 2d-2d$_{edge}$ and 3d-3d ------------------ $\gamma$=12.5 meV
  - 0d-1d, 1d-2d and 2d-3d ------------------------ $\gamma$=0.6 meV

The weak WKB tunneling prefactor is discussed in more detail in SI Appendix A. The broadening is the kinetic energy required to make the WKB prefactor =1. The broadening due to contacts is twice the broadening from a single contact given by Eq. **12**. For each dimensionality, the largest form of broadening will dominate.

The turn-on conductance versus overlap control voltage $V_{OL}$ can be seen in Fig. 7 for all of the cases. The initial broadened turn on is represented by the dotted lines.

The nanowire based devices are shown in Fig. 7(a). We see that introducing quantum confinement increases the conductance when the tunneling probability is low. For the edge tunneling devices and area tunneling devices shown in Fig. 7(b) and Fig 7(c) respectively, maximizing the quantum confinement on both sides of the junction results in the highest conductance. Overall, we see that using quantum confinement in the tunneling direction can give a sharper turn on and significantly increase the conductance when the tunneling probability is low.

**Conclusions**

Dimensionality significantly affects the I-V characteristics of tunneling diodes, including Backward Diodes and tunneling Field Effect Transistors. For pn-junctions, it becomes now necessary to specify the dimensionality of the p-region, and the dimensionality of the n-region.

**Acknowledgment**

This work was supported by the Center for Energy Efficient Electronics Sciences, which receives support from the National Science Foundation (NSF award number ECCS-0939514).

# Supplementary Information for "Pronounced Effect of pn-Junction Dimensionality on Tunnel Switch Threshold Shape"

**SI Appendix A: Transfer Matrix Element Derivation**

In our derivation of the tunnel matrix element by the transfer Hamiltonian method we will consider 3d-3d case as shown in Fig. S1(a). The method for the other reduced dimensionality cases is very similar and we will note some of the changes that would be necessary for those cases as we go through the derivation.

First we consider a simple Type III junction band diagram as shown in Fig. S1(b). The total Hamiltonian H, is illustrated in Fig. S1(b). The incomplete initial Hamiltonian, $H_i$, on the left is in Fig. S1(c), and the incomplete final state Hamiltonian $H_f$ on the right is in Fig. S1(d). For the cases in Figs. S1(c-d), the incomplete Hamiltonians lead to their own stationary Schrodinger's equations: $H_i|\Psi_i\rangle=E_i|\Psi_i\rangle$ and $H_f|\Psi_f\rangle=E_f|\Psi_f\rangle$ respectively. The subscript 'i' represents the initial electron in the valence band and the subscript 'f' represents the final electron in the conduction band.

In the true full Hamiltonian, H, a valence band electron on the left decays exponentially into the barrier, and tunnels to the conduction band on the right. The perturbation Hamiltonian with respect to the starting Hamiltonian is therefore $H'=H-H_i$. The Fermi's Golden Rule transition rate for an electron in the valence band on the left, tunneling to the conduction band on the right, is:

$$R_{if} = \frac{2\pi}{\hbar}\left|\left\langle \psi_f | H' | \psi_i \right\rangle\right|^2 \frac{dN}{dE} = \frac{2\pi}{\hbar}\left|\left\langle \psi_f | H - H_i | \psi_i \right\rangle\right|^2 \frac{dN}{dE} = \frac{2\pi}{\hbar}\left|\left\langle \psi_f | H - E_i | \psi_i \right\rangle\right|^2 \frac{dN}{dE} \qquad \textbf{[S.1]}$$

where we used the fact that $H_i|\Psi_i\rangle=E_i|\Psi_i\rangle$, and $dN/dE$ represents the density of final states.

The exact Hamiltonian, in Fig. S1(b) naturally divides into three regions. For z<0 the system resembles $H_i$, whose eigenstates are in the valence band on the left. For $0<z<W_B$, there is a barrier which the electron must tunnel through, and for $z>W_B$ the system resembles $H_f$ with eigenstates in the conduction band on the right. $\Psi_i$ is a free particle in the valence band and the exponential decay can be modeled by the WKB approximation. For convenience we segregate the problem into halves, picking a surface somewhere in the barrier so that we can divide the junction into a left half and a right half. For simplicity we choose the dividing plane to be at $W_B/2$ as shown in Fig. S1(b).

Since $(H_i-E_i)|\Psi_i\rangle=0$ everywhere, and $H\equiv H_i$ in the left half space, then $(H-E_i)|\Psi_i\rangle=0$, in the left half-space; $z<W_B/2$. Likewise, since $(H_f-E_f)|\Psi_f\rangle=0$ everywhere, and $H\equiv H_f$ in the right half-space, $(H-E_f)|\Psi_f\rangle=0$ in the right half-space; $z>W_B/2$.

Following refs. (1)&(2), the matrix element, $M_{fi} = \int_{-\infty}^{\infty} d^3r\, \psi_f^* \left(H - E_i\right)\psi_i$ can be simplified by recognizing that the integral is certainly zero for $z < W_B/2$ and by subtracting $0 = \left[\psi_i^* (H - E_f)\psi_f\right]^*$ for $z > W_B/2$. Further simplification arises when we express the Hamiltonian in the standard format:

$$H = -\frac{\hbar^2 \nabla^2}{2m} + V(r) \qquad \textbf{[S.2]}$$

where V(r) describes the entire potential of the junction. By substituting this into $M_{fi}$:

$$M_{fi} = \int_{Z>W_B/2}^{\infty} d^3r \left[\psi_f^* \left(H - E_i\right)\psi_i - \psi_i (H - E_f)\psi_f^*\right] \qquad \textbf{[S.3]}$$

and using both energy conservation, $E_i = E_f$, and the cancellation of terms involving V(r), we will be left with:

$$\begin{aligned} M_{fi} &= \frac{-\hbar^2}{2m}\int_{Z>W_B/2} d^3r \times \left(\psi_f^* \nabla^2 \psi_i - \psi_i \nabla^2 \psi_f^*\right) \\ &= \frac{-\hbar^2}{2m}\int_{Z>W_B/2} d^3r \times \nabla \cdot \left(\psi_f^* \nabla \psi_i - \psi_i \nabla \psi_f^*\right) \end{aligned} \qquad \textbf{[S.4]}$$

Now we use Gauss's law to express the matrix element as:

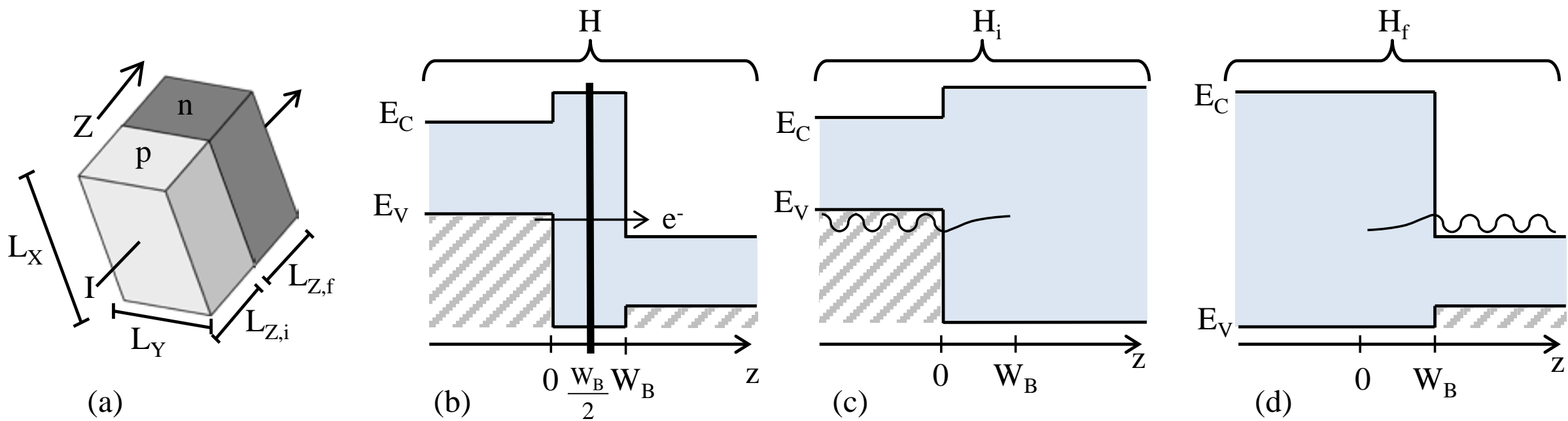

Fig. S1: (a) The 3d-3d junction that is modelled. (b) The exact total Hamiltonian $H$. (c) The incomplete Hamiltonian $H_i$ whose eigenstate represents the initial valence electron of energy $E_i$. (d) The incomplete Hamiltonian $H_f$ whose eigenstate represents the final conduction band electron of energy $E_f$.

$$M_{fi} = \hbar i \int_{Z=L/2} \vec{G}_{fi} \cdot d\vec{S} \qquad \textbf{[S.5]}$$

$$\text{with } \vec{G}_{fi} \equiv \frac{i\hbar}{2m}\left(\psi_f^* \nabla \psi_i - \psi_i \nabla \psi_f^*\right) \qquad \textbf{[S.6]}$$

Thus the matrix element is expressed as a surface integral of $\vec{G}_{fi}$ which is nonzero only at the $z=W_B/2$ surface.

### *WKB Wavefunction*

To determine the tunneling matrix element, Eq. **S.5** in the case of 3d-3d bulk tunneling we must first write down $\Psi_i$ and $\Psi_f$ in order to evaluate $G_{fi}$. Within the effective mass approximation, we can use the WKB approximation to write down the wave functions. In the low energy limit, the WKB approximation breaks down as we will see in the next section. We neglect the underlying Bloch functions, but for a more complete treatment see ref. (3). We also assume that most of the probability density is outside of the barrier region and so the barrier region can be neglected when calculating the normalization constant. The normalized WKB wavefunction becomes:

$$\psi_i = \sqrt{\frac{2k_{z,i}}{L_x L_y L_{z,i}}} \times \exp\left(ik_{x,i}x + ik_{y,i}y\right) \times \frac{1}{\sqrt{k_z(z)}} \times \begin{cases} \sin\left(\int_z^0 k(z') \times dz' + \frac{\pi}{4}\right), & z < 0 \\ \frac{1}{2}\exp\left(-\int_0^z k(z') \times dz'\right), & z \geq 0 \end{cases} \qquad \textbf{[S.7a]}$$

$$\psi_f = \sqrt{\frac{2k_{z,f}}{L_x L_y L_{z,f}}} \times \exp\left(ik_{x,f}x + ik_{y,f}y\right) \times \frac{1}{\sqrt{k_z(z)}} \times \begin{cases} \sin\left(\int_{W_B}^z k(z') \times dz' + \frac{\pi}{4}\right), & z > W_B \\ \frac{1}{2}\exp\left(-\int_z^{W_B} k(z') \times dz'\right), & z \leq W_B \end{cases} \qquad \textbf{[S.7b]}$$

In these equations $k_{\alpha,i}$ and $k_{\alpha,f}$ are the α-component of the k-vector in the initial and final states respectively. $k_z(z)$ is the spatially dependent value of $k_z$ that varies within the barrier. $L_x$, $L_y$, $L_{z,i}$, and $L_{z,f}$ are the dimensions of the device as shown in Fig. S1(a). $L_{z,i}$ represents the length of the left half of the device for $z < 0$. $L_{z,f}$ represents the length of the right half of the device for $z > W_B$. Plugging these wavefunctions into $\vec{G}_{fi}$ and evaluating it at $z=W_B/2$ gives:

$$G_{fi,\hat{Z}} = -\sqrt{\frac{k_{z,f}k_{z,i}}{L_{z,f}L_{z,i}}} \frac{i\hbar}{2mL_xL_y} \exp\left(i\Delta k_x x + i\Delta k_y y\right) \times \exp\left(-\int_0^{W_B} k_z dz\right) \qquad \textbf{[S.8]}$$

where $\Delta k_x = (k_{x,i} - k_{x,f})$ and $\Delta k_y = (k_{y,i} - k_{y,f})$. Using this and evaluating the expression for the matrix element we get:

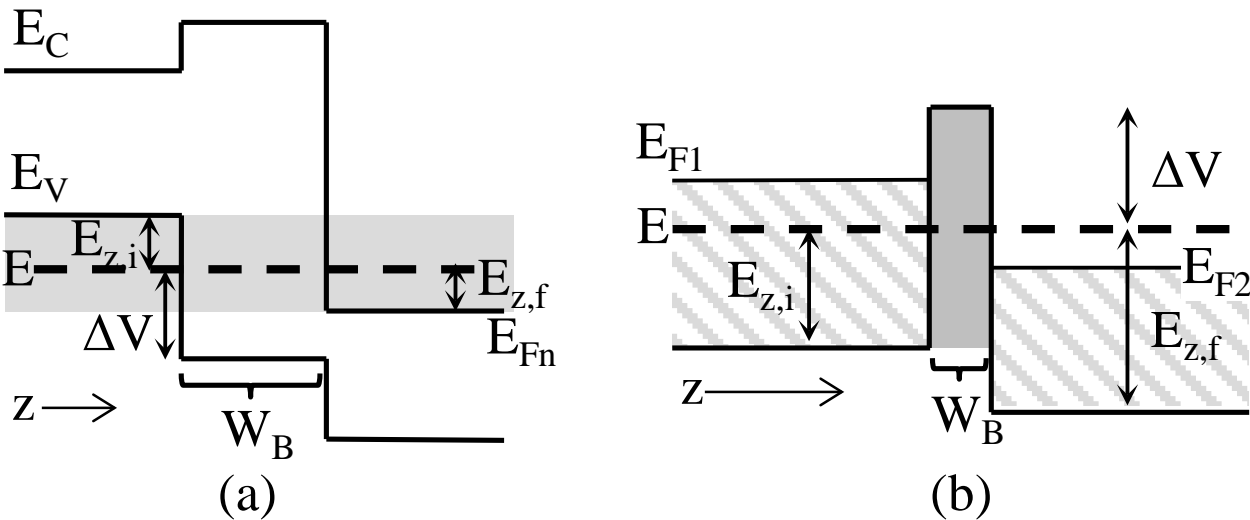

Fig. S2: (a) A rectangular band to band tunneling barrier. (b) A rectangular single band tunneling barrier.

$$M_{fi} = \frac{\hbar^2}{2m}\sqrt{\frac{k_{z,f}k_{z,i}}{L_{z,f}L_{z,i}}} \times \exp\left(-\int_0^{W_B} k_z dz\right) \times \delta_{k_{x,i},k_{x,f}}\,\delta_{k_{y,i},k_{y,f}} \quad \textbf{[S.9]}$$

The kronecker deltas represent the conservation of transverse momentum and show that the conservation is a natural result of calculating the matrix element. For the case of incomplete conservation of momentum, the kronecker deltas will be replaced by the actual surface integral in Eq. **S.5**. At this point, we replace the WKB integral $\exp\left(-2\int_0^{W_B} k_z dz\right)$ with $\top$. In practice, the tunneling probability may vary from the simple WKB integral in some situations (4) and so $\top$ can be generalized to reflect those changes. Thus the matrix element is given by:

$$M_{fi} = \frac{\hbar^2}{2m}\sqrt{\frac{k_{z,f}k_{z,i}}{L_{z,f}L_{z,i}}} \times \sqrt{\top} \times \delta_{k_{x,i},k_{x,f}}\,\delta_{k_{y,i},k_{y,f}} \quad \textbf{[S.10]}$$

Interestingly, this expression is also valid for all of the reduced dimensionality cases, we just need to sum over fewer k-states.

For the reduced dimensionality cases we can use $k_z=\pi/L_z$ and $E_z = \hbar^2 k_z^2/2m^*$ to further simplify the matrix element. For 0d-1d we get the following matrix element:

$$M_{fi,0d-1d} = \sqrt{\frac{E_{z,i}}{\pi} \times \left(\frac{\hbar^2}{2m}\frac{k_{z,f}}{L_{z,f}}\right) \times \top} \quad \textbf{[S.11]}$$

For 0d-0d both sides of the junction are confined which gives:

$$M_{fi,0d-0d} = \frac{1}{\pi}\sqrt{E_{z,i} \times E_{z,f} \times \top} \quad \textbf{[S.12]}$$

***Rectangular Barrier Wavefunction***

Instead of using the WKB approximation, we can also use the wavefunctions that correspond to a rectangular barrier as shown in Fig S2(a). In this case, we can write down the exact wavefunction, but we cannot account for exact barrier shape. This will allow us to see the low energy dependence of the tunneling probability where the WKB approximation fails.

The normalized wavefunctions for a rectangular barrier are:

$$\psi_i = \sqrt{\frac{2}{L_x L_y L_{z,i}}} \times \exp\left(ik_{x,i}x + ik_{y,i}y\right) \times \begin{cases} \cos\left(k_{z,i}(z + L_{z,i}/2)\right), & z < 0 \\ \dfrac{1}{\sqrt{1+\left(\kappa/k_{z,i}\right)^2}} \times \exp\left(-\kappa z\right), & z \ge 0 \end{cases} \quad \textbf{[S.13a]}$$

$$\psi_f = \sqrt{\frac{2}{L_x L_y L_{z,f}}} \times \exp\left(ik_{x,f}x + ik_{y,f}y\right) \times \begin{cases} \cos\left(k_{z,f}(z - L_{z,f}/2 - W_B)\right), & z > W_B \\ \dfrac{1}{\sqrt{1+\left(\kappa/k_{z,f}\right)^2}} \exp\left(\kappa (z - W_B)\right), & z \le W_B \end{cases} \quad \textbf{[S.13b]}$$

κ is the wavevector in the barrier and is given by: $\kappa = \sqrt{2m\Delta V} / \hbar$ . $\Delta V$ is the barrier height relative to the energy as shown in Fig. S2(a). Plugging these wavefunctions into **S.5** and evaluating the matrix element gives:

$$M_{fi} = \frac{\hbar^2}{2m}\sqrt{\frac{k_{z,f}k_{z,i}}{L_{z,f}L_{z,i}}} \times \left( \frac{4\kappa\sqrt{k_{z,i}k_{z,f}}}{\sqrt{\left(k_{z,i}^2 + \kappa^2\right)\left(k_{z,f}^2 + \kappa^2\right)}} \right) \exp\left(-\kappa W_B\right) \times \delta_{k_{x,i},k_{x,f}} \delta_{k_{y,i},k_{y,f}} \quad \textbf{[S.14]}$$

Comparing this with Eq. **S.10** we find that:

$$T = \left( \frac{16\kappa^2 k_{Z,i}k_{Z,f}}{\left(k_{z,i}^2 + \kappa^2\right)\left(k_{z,f}^2 + \kappa^2\right)} \right) \exp\left(-2\kappa W_B\right) = \frac{16\Delta V\sqrt{E_{z,i}E_{z,f}}}{\left(E_{z,i} + \Delta V\right)\left(E_{z,f} + \Delta V\right)} \exp\left(-2\kappa W_B\right) \quad \textbf{[S.15]}$$

At small energies, $E<<\Delta V$, we get:

$$T \approx \frac{16\sqrt{E_{z,i}E_{z,f}}}{\Delta V} \exp\left(-2\kappa W_B\right) \quad \textbf{[S.16]}$$

Thus we see that there is an energy dependent pre-factor to the WKB exponential. As the energy goes to zero, the tunneling probability goes to zero. Near turn on, this can have a significant effect.

When there is no confinement in the tunneling direction in the 1d-1$d_{end}$, 2d-2$d_{edge}$ and 3d-3d cases, the available kinetic energy, $E_z$, is limited by $V_{OL}$ and this gives a voltage dependent turn on. The tunneling probability, Eq. **S.16**, will be maximized when $E_{z,i} = E_{z,f} = qV_{OL}/2$:

$$T_{max} \approx \frac{8V_{OL}}{\Delta V} \exp\left(-2\kappa W_B\right) \quad \textbf{[S.17]}$$

This means that the tunneling probability will be linearly proportional to $V_{OL}$ at turn-on and a finite $V_{OL}$ of ΔV/8 is needed for the pre-factor to reach 1. Once the prefactor is 1, we assume the WKB approximation is valid.

For the 0d-1d, 1d-2d and 2d-3d cases, one side of the junction is confined and so we need to use Eq. **S.16** with $E_{z,f} = qV_{OL}/2$ . The turn on will be broadened until the prefactor is equal to 1 at which point the WKB approximation is valid:

$$16\sqrt{E_{z,i}qV_{OL}/2} \Big/ \Delta V = 1 \quad \textbf{[S.18]}$$

***One Band Rectangular Barrier***

To verify the matrix element derivation, the tunnel probability derived from computing the matrix element can be compared to the 1 band tunneling probability through a single barrier as shown in Fig. 3(b). The 1 band tunneling probability through a rectangular barrier can be found be matching boundary conditions using propagation matrices (5) and is given by:

$$T = \frac{1}{\dfrac{\left(\sqrt{E_{z,i}} + \sqrt{E_{z,f}}\right)^2}{4\sqrt{E_{z,i}E_{z,f}}} + \dfrac{\left(E_{z,i} + \Delta V\right)\left(E_{z,f} + \Delta V\right)}{4\Delta V\sqrt{E_{z,i}E_{z,f}}} \sinh^2(\kappa W_B)} \quad \textbf{[S.19]}$$

We considered the situation where the initial and final energy, $E_{z,i}$ and $E_{z,f}$ respectively, are different as shown in Fig. S2(b). The barrier height relative to the tunneling energy, $E$, is given by $\Delta V$. The barrier width is $W_B$. The wavevector in the tunneling barrier is given by: $\kappa = \sqrt{2m\Delta V} \big/ \hbar$ . For a typical barrier the sinh term will be large and so we get:

$$T \approx \frac{16\Delta V\sqrt{E_{z,i}E_{z,f}}}{\left(E_{z,i} + \Delta V\right)\left(E_{z,f} + \Delta V\right)} \exp\left(-2\kappa W_B\right) \quad \textbf{[S.20]}$$

This is identical to Eq. **S.15** and thus we verified that computing the matrix element gives the same tunneling probability as directly calculating it by matching boundary conditions. Since we are using the effective mass approximation, same band tunneling and band to band tunneling give the same result so long as the correct definition of $E_z$ is used.

**SI Appendix B: Using the Transfer Matrix Element to Derive Current**

In SI Appendix A we found the matrix element that can be used with Fermi's golden rule. Using this, we can now find the current in any of the different cases. To aid in correctly counting the number of states, we use the delta function version of Fermi's golden rule. The transition rate between two states is:

$$R_{if} = \frac{2\pi}{\hbar}\left|\langle \psi_f / H - E_i / \psi_i \rangle\right|^2 \delta(E_i - E_f) \qquad \textbf{[S.21]}$$

We convert the transition rate to a tunneling current by multiplying the rate by the electron charge, summing over initial and final states, and multiplying by the Fermi-Dirac occupation probabilities.

$$J_{Tunnel} = 2q \sum_{k_i,k_f} R_{if} f_v(1-f_c) - R_{fi} f_c(1-f_v) \qquad \textbf{[S.22a]}$$

$$= 2q \sum_{k_i,k_f} R_{fi}(f_c - f_v) \qquad \textbf{[S.22b]}$$

$$= \frac{4\pi q}{\hbar} \sum_{k_i,k_f} \left|M_{fi}\right|^2 \delta(E_i - E_f)(f_c - f_v) \qquad \textbf{[S.22c]}$$

Where

$$f_v = \frac{1}{\exp\left[(E_i - F_p)/k_b T\right] + 1} \qquad \textbf{[S.23a]}$$

$$f_c = \frac{1}{\exp\left[(E_f - F_n)/k_b T\right] + 1} \qquad \textbf{[S.23b]}$$

$F_n$ and $F_p$ are the quasi Fermi levels for electrons and holes respectively. Plugging the matrix element Eq. **S.10** into Eq. **S.22c** for tunneling current gives:

$$I_{Tunnel} = \frac{\pi q \hbar^3}{m^2} \sum_{k_i,k_f} \frac{k_{z,i} k_{z,f}}{L_{z,i} L_{z,f}} (f_c - f_v) \times \top \times \delta_{k_{x,i},k_{x,f}} \delta_{k_{y,i},k_{y,f}} \delta(E_i - E_f) \qquad \textbf{[S.24a]}$$

$$I_{Tunnel} = \frac{8q}{h} \sum_{k_i,k_f} \left( \frac{\hbar^2 k_{z,i}}{2m} \frac{\pi}{L_{z,i}} \right) \times \left( \frac{\hbar^2 k_{z,f}}{2m} \frac{\pi}{L_{z,f}} \right) \times (f_c - f_v) \times \top \times \delta_{k_{x,i},k_{x,f}} \delta_{k_{y,i},k_{y,f}} \delta(E_i - E_f) \qquad \textbf{[S.24b]}$$

Interestingly, this expression is also valid for all of the reduced dimensionality cases, we just need to sum over fewer k-states.

***3d-3d Bulk Junction***

For 3d to 3d bulk we break the sums up into a transverse ($k_t$) and z component ($k_z$) and then we convert the sums over $k_z$ and into integrals of $E_z$ using:

$$\sum_{k_z} = \frac{L_z m}{\pi \hbar^2} \int \frac{dE_z}{k_z} \qquad \textbf{[S.25]}$$

Plugging this into Eq. **S.24a** gives:

$$I_{3d-3d} = \frac{q}{\pi \hbar} \sum_{k_t} \int dE_{z,i} \int dE_{z,f} \times (f_C - f_V) \times \top \times \delta(E_i - E_f) \qquad \textbf{[S.26]}$$

The sum over $k_t$ can be converted to an energy integral using the following:

$$\sum_{k_t} = \frac{m}{2\pi \hbar^2} \int dE_t \qquad \textbf{[S.27]}$$

Plugging this into Eq. **S.26** and eliminating one of the energy integrals by evaluating the delta function gives:

$$I_{3d-3d} = \frac{qmA}{2\pi^2 \hbar^3} \int dE_t \int dE_{z,i} (f_C - f_V) \times \top \qquad \textbf{[S.28]}$$

Next, we can then convert the integrals over $E_{z,i}$ to an integral over total energy, E, by a change of variables and change the order of the integrals:

$$I_{3d-3d} = \frac{qmA}{2\pi^2\hbar^3}\int_0^{qV_{OL}} dE_i \int_0^{\min(E,V_{OL}-E)} dE_t \left(f_C - f_V\right)\times \top \qquad \textbf{[S.29]}$$

Here we take the zero of energy to be at the conduction band edge on the n-side. The transverse energy can be no more than the total energy on either side of the junction. For reduced dimensionalities we will be summing over fewer k-states and so there may be only one or even no integrals.

If we assume a full valence band and empty conduction band we can set $\left(f_C - f_V\right) = 1$ and evaluate the integrals to recover Eq. **8**

$$I_{3d-3d} = \frac{1}{2}\left(\frac{Am^*}{2\pi\hbar^2}\times\frac{qV_{OL}}{2}\right)\times\frac{2q^2}{h}V_{OL}\times\langle\top\rangle \qquad \textbf{[S.30]}$$

Thus we have finally recovered the equation for 3d-3d bulk tunneling current. We can also consider small biases less than $k_BT$ by setting $\left(f_C - f_V\right) \approx qV_{SD}/(4k_BT)$.

### *2d-3d Bulk Junction*

Next, we consider the 2d-3d case to demonstrate the general applicability of Eq. **S.24**. In this case we sum over the transverse states ($k_t$) and only the final $k_z$ states. After converting the sums to energy integrals and evaluating the delta function we get:

$$I_{2d-3d} = \frac{Am}{2\pi\hbar^2}\times\frac{4q}{h}\times\int_0^{qV_{OL}/2} dE_t\left(f_C - f_V\right)\times\top \qquad \textbf{[S.31]}$$

Assuming a full valence band and empty conduction we recover Eq. **13**

$$I_{2d\text{-}3d} = \left(\frac{Am}{2\pi\hbar^2}\times\frac{qV_{OL}}{2}\right)\times\left(\frac{4q}{h}\times E_z\times\langle\top\rangle\right) \qquad \textbf{[S.32]}$$

### *2d-2d$_{face}$ Junction*

For the 2d-2d face junction we have only a single $k_{z,i}$ and a single $k_{z,f}$ corresponding to the confinement energies. Starting from Eq. **S.24b** and evaluating the kronecker delta function gives:

$$I_{2d-2d,face} = \frac{8q}{h}\sum_{k_t} E_{z,i}\times E_{z.f}\times\left(f_c - f_v\right)\times\top\times\delta(E_i - E_f) \qquad \textbf{[S.33]}$$

Next we convert the sum over tranverse states into an integral over energy using Eq. **S.27**. This gives Eq. **23**:

$$I_{2d-2d,face} = 2q\times\frac{Am}{\pi^2\hbar^3}\int E_{z,i}\times E_{z,f}\times\langle\top\rangle\times\delta(E_i - E_f)dE_t \qquad \textbf{[S.34]}$$

Finally evaluating the integral recovers Eq. **24**:

$$I_{2d-2d,face} = \frac{qmA}{\pi^2\hbar^3}\times E_{z,i}\times E_{z,f}\times\langle\top\rangle \qquad \textbf{[S.35]}$$

Similarly, we can derive the current for any of the cases using Eq. **S.24**